# Supervision by Denoising

Sean I. Young ⬡, Adrian V. Dalca ⬡, Enzo Ferrante ⬡, Polina Golland ⬡,
Christopher A. Metzler ⬡, Bruce Fischl ⬡, and Juan Eugenio Iglesias ⬡

**Abstract**—Learning-based image reconstruction models, such as those based on the U-Net, require a large set of labeled images if good generalization is to be guaranteed. In some imaging domains, however, labeled data with pixel- or voxel-level label accuracy are scarce due to the cost of acquiring them. This problem is exacerbated further in domains like medical imaging, where there is no single ground truth label, resulting in large amounts of repeat variability in the labels. Therefore, training reconstruction networks to generalize better by learning from both labeled and unlabeled examples (called semi-supervised learning) is problem of practical and theoretical interest. However, traditional semi-supervised learning methods for image reconstruction often necessitate handcrafting a differentiable regularizer specific to some given imaging problem, which can be extremely time-consuming. In this work, we propose "supervision by denoising" (SUD), a framework to supervise reconstruction models using their own denoised output as labels. SUD unifies stochastic averaging and spatial denoising techniques under a spatio-temporal denoising framework and alternates denoising and model weight update steps in an optimization framework for semi-supervision. As example applications, we apply SUD to two problems from biomedical imaging—anatomical brain reconstruction (3D) and cortical parcellation (2D)—to demonstrate a significant improvement in reconstruction over supervised-only and ensembling baselines. Our code available at https://github.com/seannz/sud.

✦

## 1 INTRODUCTION

L EARNING-BASED IMAGE RECONSTRUCTION MODELS have become indispensable for solving a vast array of inverse problems in imaging and vision, from low-level ones which involve image denoising and reconstruction [1], [2] to high-level ones that necessitate pixel-level semantic segmentation [3], [4]. While reconstruction models such as those based on U-Net [5] typically outperform handcrafted models in many imaging problems, they can involve millions of parameters and, as a result, have a tendency to overfit training data and generalize poorly to previously unseen images at test time— a problem also exacerbated by distribution shift [6]. Such an issue is especially noticeable in the reconstruction of medical images, where labeled training images are scarce, produced under one of many different imaging (contrast) settings and contain label noise due to the variability in the labeling even across domain experts [7]. Therefore, training reconstruction networks to generalize better by learning from both labeled and unlabeled examples (called semi-supervised learning) is a problem of practical and theoretical interest for improving image reconstruction and other downstream tasks [8], [9].

- S I Young, A V Dalca, and J E Iglesias are with the Martinos Center for Biomedical Imaging, Harvard Medical School, Boston, MA, USA, and the Computer Science and Artificial Intelligence Lab, MIT, Cambridge, MA, USA. E-mail: siyoung@mit.edu, adalca@mit.edu, jei@mit.edu.
- E Ferrante was with the Martinos Center for Biomedical Imaging, Harvard Medical School, Boston, MA, USA. He is now with CONICET, Santa Fe, Argentina. E-mail: eferrante@sinc.unl.edu.ar.
- P Golland is with the Computer Science and Artificial Intelligence Lab (CSAIL), MIT, Cambridge, MA, USA. E-mail: polina@csail.mit.edu.
- C A Metzler is with the Department of Computer Science, University of Maryland, College Park, MD, USA. E-mail: metzler@umd.edu.
- B Fischl is with the Martinos Center for Biomedical Imaging, Harvard Medical School, Boston, MA, USA. E-mail: bfischl@mgh.harvard.edu.

Broadly speaking, we can improve the generalization of a reconstruction network by altering aspects of its architecture [10], [11], training data [12], [13], training objectives [14], [15] or optimization strategies used for training [16], [17]. Out of these, adding a spatial regularizer in the training objective of the reconstruction network has proved extremely useful for imposing topological or spatial priors on the reconstruction [18], [19] and semi-supervised learning (SSL). SSL methods based on regularization suffer neither from limited diversity of augmented data nor domain gaps resulting from training on synthetic data [20], [21]. Moreover, they explicitly specify the spatial regularity needed for a particular task, in contrast to augmentation [22] and self-supervision [23] techniques.

Despite the benefits, incorporating spatial regularizers to facilitate semi-supervision has been difficult due to the need to handcraft a differentiable regularizer for a given problem [18], [24], which can be extremely time-consuming. Classical differentiable regularizers [25], [26], by contrast, are agnostic to the task and produce sub-optimal outcomes. One solution to overcome the weaknesses of conventional regularizers is to construct them from denoisers, which would allow us to convert domain-specific, learned or even non-differentiable denoisers to regularizers. This paradigm is also followed by frameworks such as "regularization by denoising" [27], [28] and "plug-and-play priors" [29], [30], which are applied to solve various inverse problems in imaging [31], [32]. Given the fundamental role of denoising in imaging tasks, it is not so surprising that supervising reconstruction networks can benefit from denoising as well. In fact, stochastic averaging methods used in learned image reconstruction [33], [34] can be thought of as "regularization by temporal denoising" and this suggests regularization by denoising in both space and time (across models) can lead to better semi-supervision.

In this work, we propose "supervision by denoising", or SUD, a framework that allows supervision of reconstruction models using their denoised output as targets. SUD unifies spatial filtering [27], [30] and stochastic averaging [23], [35] techniques under one spatio-temporal denoising framework and alternates denoising and network weight update in an optimization framework for principled supervision. While SUD accommodates any learned and even non-differentiable reconstruction denoiser, simply just training one alleviates the need to find a denoiser suited for a given task. We learn reconstruction denoisers by training them self-supervised on reconstruction examples (segmentations or images) already available. Fig. 1 summarizes the SUD framework with data augmentation and dropout, noting that the denoising is both spatial and temporal (across training steps) as shown by (a) and (b). In Fig. 2, we show one application of SUD (cortical parcellation) and demonstrate that it significantly improves supervision of a well-performing baseline network [36]. We note that denoising reconstructions post hoc produces poor outcomes whereas naively alternating gradient descent and denoising renders training divergent.

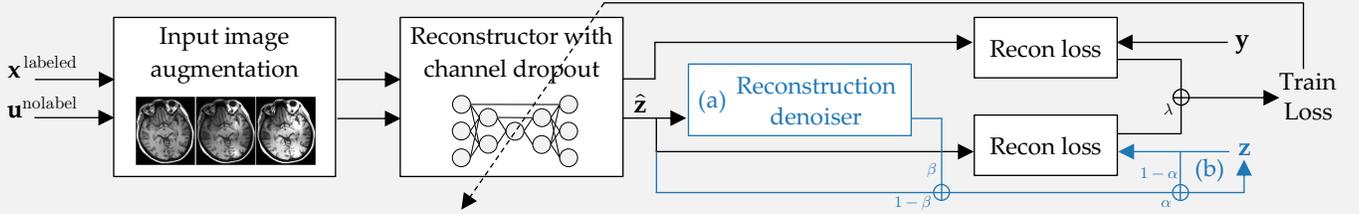

Fig. 1. Supervision by Denoising. We train a reconstruction network on unlabeled data $\mathbf{u}^{\text{nolabel}}$. Reconstructor output $\hat{\mathbf{z}}$ is denoised spatially using a spatial denoiser (a) then stochastic averaging (b) to produce pseudo-target $\mathbf{z}$. The reconstruction loss between $\hat{\mathbf{z}}$ and $\mathbf{z}$ is scaled and added to that of labeled pair ($\mathbf{x}^{\text{labeled}}$, $\mathbf{y}$) and the combined used to update reconstructor weights. The denoising pipeline is highlighted in blue. See Algorithm 1.

Our contributions are as follows. First, we derive SUD by reinterpreting a semi-supervised image reconstruction task as an optimization problem that can be solved by alternating denoising and stochasticgradient descent steps. Second, we apply SUD on two tasks from medical imaging (2D cortical parcellation and 3D brain reconstruction) and demonstrate a substantial improvement in the reconstruction quality over supervised and stochastic averaging baselines, especially in the barely supervised (with few labeled examples) scenario.

## 2   Related Work

Regularization by denoising (RED) [27] is a classical image reconstruction method that uses denoisers to regularize an image. By contrast, our work is a semi-supervised learning method that uses denoisers to regularize learning. We briefly review RED [27] and non-denoising-based semi-supervision techniques for image reconstruction as a leadup to SUD.

### 2.1   Regularization by Denoising

Many inverse problems in imaging and vision are inherently ill-conditioned or even ill-posed and require an assumption of "regularity" to recover a plausible solution. Regularizing inverse problems with traditional techniques often involves a domain-agnostic regularity term such as the total variation semi-norm and its generalizations [25], [26] or graph-based (non-local) regularizers [37], [38], both of which are usually chosen out of modeling or optimization convenience, rather than to reflect a true notion of regularity in a chosen domain.

Among these regularizers, the bilateral regularizer [39] is shown to be effective for solving computer vision problems such as image refocus [40], optical flow estimation [37], [41] and image superresolution [42]. Bilateral regularization [39] essentially expresses regularity of an image as its $\ell^2$ distance from a bilateral-filtered version of the same image to define the prior in an anisotropic diffusion space [43]. Other classic edge-preserving filters that can be plugged into the bilateral solver include non-local means denoising [44] and trilateral [45] filters, both of which can be implemented efficiently.

The "regularization by denoising" (RED) framework [27] can be seen as extending bilateral regularization, expressing regularizers in terms of arbitrary image denoisers. Reehorst and Schniter [28] clarify the conditions on such denoisers to guarantee convergence while [46] demonstrate the utility of learned denoisers for regularization. Closely related to RED is the older "plug-and-play priors" (P3) method [29], which uses a denoising step in place of the denoising optimzation in RED. While SUD also uses an optimization-free denoising strategy to accelerate training, this strategy is derived from spectral filter theory [47] to ensure the denoising step more faithfully approximates this optimization. We stress neither RED nor P3 addresses supervision as in our work.

Recently proposed image denoising techniques typically dispense with models of diffusion [43], and learn from data a convolutional neural network which map noisy images to noise-free ones directly. Denoising based on U-Net has been applied to a wide variety of image types, including medical [48], [49], satellite [50] and natural [51] ones. Such flexibility of U-Net has also been applied to denoising reconstructions [52], [53], [54], obviating the need to denoise via discrete filters [55], [56], which are computationally expensive. SUD leverages such denoisers to supervise image reconstruction.

### 2.2   Semi-supervised Learning

In the machine learning literature, semi-supervised learning (SSL) is a family of approaches that learn using both labeled and unlabeled data, allowing learning algorithms to exploit abundant unlabeled data to enhance a smaller set of labeled examples. Currently, a large number of well-performing SSL methods involve adding a regularity loss on unlabeled data to the training objective, leaving all other aspects of training unchanged [57]. Broadly speaking, we can classify such SSL methods as either stochastic (used in both classification and reconstruction tasks) or spatial (used only in reconstruction) since classification output do not have spatial dimensions.

Stochastic ensembling techniques [23], [33], [34], [35] add a regularity term to the training objective to ensure that the network output for a given image does not deviate too much from its previous values. Used for classification tasks, the $\Pi$ model [58] is a simple technique that penalizes deviation of the current output from its last value, essentially treating the last output as a soft or pseudo-target. Temporal ensembling

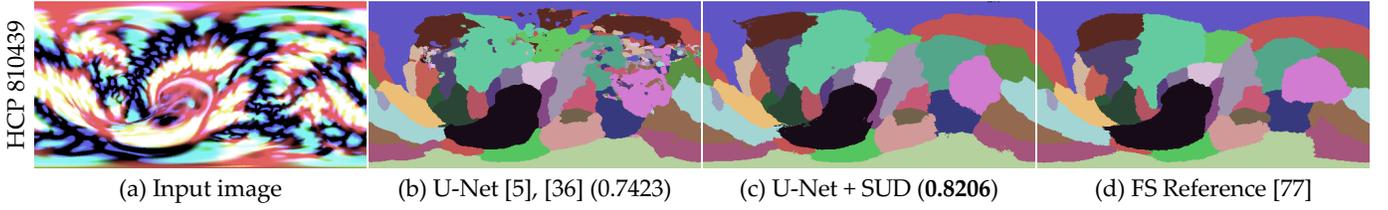

HCP 810439

(a) Input image     (b) U-Net [5], [36] (0.7423)     (c) U-Net + SUD (**0.8206**)     (d) FS Reference [77]

Fig. 2. SUD applied to "cortical parcellation". Validation results shown with Dice in parentheses. One labeled image used for training. Input (a) is a three-channel image consisting of (sulcal map, white matter curvature, inflated surface curvature). Baseline U-Net produces noisy and anatomically implausible reconstruction (b) while training it additionally under SUD produces a more plausible one (c) that is closer to the reference (d).

(TE) [23] improves upon the Π model by averaging previous network output to obtain pseudo-targets. The mean teacher model [35] is a variant of [23] which uses the output from a temporally averaged network as pseudo-targets. Very recent extensions to TE are based on aggressive data augmentation and postprocessing of the network output [57], [59], multiple teacher models to generate the pseudo-targets [60], [61], and transformation-consistency [34], [35] particularly in the case of reconstruction tasks. Later we show TE can be interpreted as a form of "supervision by temporal denoising" and extend the denoising to spatial dimensions to improve supervision of reconstruction networks. In the process, we also reveal the implicit regularized training objective optimized by TE.

Spatial regularization has been employed extensively in image reconstruction problems to impose topological and spatial constraints on reconstruction. These regularizers can also be used as an unsupervised loss term on unlabeled data to facilitate semi-supervised learning. Star-shape regularizer [24] penalizes islands and holes in the reconstruction, while the topological loss of Clough *et al.* [18] penalizes violations in the topology of the reconstruction, such as the number of holes, handles, and connected components. The regularizer of [62] uses a GAN to penalize reconstructed pixels deemed fake. Zhou *et al.* [63] penalize the size of structures deviating too much from a prior size distribution whereas those in [54] and [52], deviating from a prior shape distribution. SUD can also be thought of as regularizing shape but unlike [52], [54] applies this regularization towards SSL and operates within a principled optimization-based framework.

## 3 MATHEMATICAL FRAMEWORK

In essence, SUD leverages a reconstruction denoiser to train the main reconstruction network on unlabeled data; refer to Fig. 1 for the end-to-end training procedure. Given a labeled image, our objective is simply to minimize the loss between the reconstruction and the corresponding true label. For an unlabeled image, however, our objective is now to minimize the loss between the reconstruction and the denoised version of it. The two losses are added and backpropagated through to update the network's parameters. This is summarized in Algorithm 1. We now derive the SUD algorithm.

### 3.1 Regularizing Reconstruction

From a learning perspective, the estimation of an unknown underlying image $\mathbf{f}(\mathbf{x})$ for input $\mathbf{x}$ involves maximizing the conditional probability of $\mathbf{f}(\mathbf{x})$ given $\mathbf{x}$. Maximum likelihood is typically used in supervised learning and essentially leads to solving an optimization problem of the form

$$\text{minimize } G(\mathbf{\Theta}) = \tfrac{1}{L}\sum_{n=1}^{L} D(\mathbf{y}^n, \mathbf{f}(\mathbf{x}^n | \mathbf{\Theta})), \quad (1)$$

in which $\mathbf{x}^n$ denotes an image, and $\mathbf{f}(\mathbf{x}^n | \mathbf{\Theta})$, the mapping of

---

**Algorithm 1.** Supervision by Denoising

1   **Input:** $\{\mathbf{x}^n, \mathbf{y}^n\}^{n=1,\dots,L}$ *(labeled),* $\{\mathbf{u}\}^{n=1,\dots,U}$ *(unlabeled),*
2   $\mathbf{f}(\cdot\,|\mathbf{\Theta}^0)$ *(reconstructor to train),* $\mathbf{a}(\cdot)$ *(trained denoiser),*
3   $\beta\,(=0.125,$ *denoiser strength),*
4   $\lambda_{\max}\,(=8,$ *self-supervision weight), $N$ (training steps),*
5   **Output:** $\mathbf{f}(\cdot\,|\mathbf{\Theta}^N)$ *(trained reconstructor)*
6   $\mathbf{z}^n =$ *Initialize all soft targets to zero*
7   **for** $n = 1, \dots, N$ **do**
8     $\alpha = 1 - (n/N), \quad \lambda = (n/N)\lambda_{\max}$   *// Sec 3.3.2*
9     $\mathbf{x}^n, \mathbf{y}^n =$ *Load nth (%L) image–label pair*
10    $\mathbf{u}^n, \mathbf{z}^n =$ *Load nth (%U) unlabeled image and soft-target*
11    $\mathbf{z}^n = \alpha\beta\mathbf{a}(\mathbf{f}(\mathbf{u}^n|\mathbf{\Theta}^n)) + \alpha(1-\beta)\mathbf{f}(\mathbf{u}^n|\mathbf{\Theta}^n) + (1-\alpha)\mathbf{z}^n$
12    $\ell = D(\mathbf{y}^n, \mathbf{f}(\mathbf{x}^n|\mathbf{\Theta})) + \lambda D(\mathbf{z}^n, \mathbf{f}(\mathbf{u}^n|\mathbf{\Theta}))$
13    $\mathbf{\Theta}^{n+1} =$ *Update $\mathbf{\Theta}^n$ by back-propagating loss $\ell$*
14   **end for**

---

image $\mathbf{x}^n$ by a network $\mathbf{f}$ parameterized by $\mathbf{\Theta}$. We denote by $D(\mathbf{y}^n, \mathbf{f}^n)$, a measure of loss between $\mathbf{f}^n = \mathbf{f}(\mathbf{x}^n|\mathbf{\Theta})$ and label $\mathbf{y}^n$. Typical loss functions $D$ for medical imaging problems include categorical cross entropy and the Dice loss.

If the training dataset contains only a small number $L$ of image–label pairs $(\mathbf{x}^n, \mathbf{y}^n)$, minimizing objective $G(\mathbf{\Theta})$ could lead to overfitting and hamper generalization of $\mathbf{f}$ to images outside the training distribution. As well as diversifying the training distribution via data augmentation, one can turn to maximum a posteriori probability (MAP) estimation, where a log-prior (or regularization term) for network parameters is introduced to the training objective:

$$F(\mathbf{\Theta}) = G(\mathbf{\Theta}) + \tfrac{1}{U}\sum_{n=1}^{U} \lambda R(\mathbf{f}(\mathbf{u}^n|\mathbf{\Theta})), \quad (2)$$

in which $\mathbf{u}^n$ denotes an image, similarly to $\mathbf{x}^n$, $R$ represents a regularizing functional, such as the topological loss [18] or the total variation semi-norm [25] and $\lambda$ controls the weight of the $R$ terms in the objective. Since images $\mathbf{u}^n$ need not be the same as images $\mathbf{x}^n$ nor annotated with labels, optimizing objective (2) can be a useful way to improve generalization if one has access to very few (say, $L$) labeled images but more ($U$) unlabeled ones, which is frequently the case in medical image reconstruction problems. In such a "semi-supervised learning" (SSL) paradigm, a suitable choice of $R$ and trade-off $\lambda$ can help to leverage both labeled and unlabeled data to improve generalization of the trained network [18], [63]. We illustrate this in Fig. 3, comparing validation accuracy with and without regularization for the cortical parcellation task.

In many cases, the regularizer $R$ used in semi-supervised reconstruction is mathematically crafted for a given problem [24], [18] and often requires significant effort to guarantee it is differentiable [18]. This work proposes to overcome such difficulties by constructing $R$ as a function of a denoiser and optimizing objective (2) by alternating the denoising and the stochastic gradient descent steps.

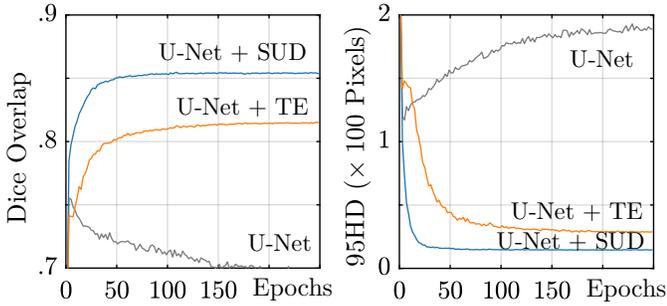

Fig. 3. Validation curves for cortical parcellation showing the effect of regularization. U-Net (nn-UNet) is trained using one labeled image and channel dropout with drop p of 0.05. Dice and 95HD worsen across epochs for U-Net but improve when regularized using SUD or temporal ensembling (TE). See Fig. 2 for visualizations of reconstructions.

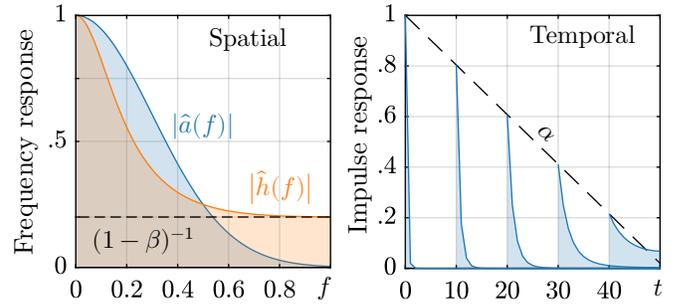

Fig. 4. Spatial (left) and temporal (right) filter responses. Optimization-based spatial filters (left, orange) add an all-pass frequency component $\beta$ while direct smoothing filters (left, blue) do not. Temporal filtering with parameter $a$ graduated across step $n$ induces time-dependent temporal impulse responses.

## 3.2 Regularization by Denoising (RED)

As a starting point, we analyze regularization by denoising (RED) [27], [28], proposed for optimizing an objective of the form $F(\mathbf{f}) = D(\mathbf{x}, \mathbf{f}) + \lambda R(\mathbf{f})$ found in many low-level vision and imaging problems. Typically, the $D$ term measures the fidelity of $\mathbf{f}$ to some given observation $\mathbf{x}$ whereas the $R$ term quantifies the regularity of $\mathbf{f}$ using such measures as the TV seminorm or the bilateral regularizer. The RED framework instead expands the regularity term into the bilinear form

$$2R(\mathbf{f}) = \langle \mathbf{f}, \mathbf{h}(\mathbf{f}) \rangle = \langle \mathbf{f}, \mathbf{f} - \mathbf{a}(\mathbf{f}) \rangle, \quad (3)$$

in which $\mathbf{h}(\mathbf{f})$ represents a non-linear Laplacian of $\mathbf{f}$, and $\mathbf{a}(\mathbf{f})$ is the corresponding smoothing operation. This allows us to regularize $\mathbf{f}$ using a non-differentiable denoising filter $\mathbf{a}$ and still express the gradient of $R$ via $R'(\mathbf{f}) = \mathbf{f} - \mathbf{a}(\mathbf{f})$, provided that the Jacobian $\mathbf{A} = J\mathbf{a}(\mathbf{f})$ is symmetric, $\mathbf{a}$ is differentiable and locally homogeneous. Regularizers contructed this way are reported to work even when the above conditions do not hold [64]. If gradient descent is used to find the optimal $\mathbf{f}$, the $n$th update step is equivalent to denoising the iterate $\mathbf{f}^n$ to obtain $\mathbf{z}^n = \mathbf{a}(\mathbf{f}^n)$ then descending along

$$\mathbf{d}^n = -\frac{\partial(D(\mathbf{x}, \mathbf{f}) + (\lambda/2)\|\mathbf{z}^n - \mathbf{f}\|_2^2)}{\partial \mathbf{f}}(\mathbf{f}^n), \quad (4)$$

noting that the derivative of $(1/2)\|\mathbf{z}^n - \mathbf{f}\|_2^2$ at $\mathbf{f}^n$ is $R'(\mathbf{f}^n)$. In this work, (stochastic) gradient descent is of special interest as it is used later to train reconstruction networks.

While RED is not designed for training networks, we can attempt to incorporate RED in semi-supervised learning by substituting (3) in objective (2). Applying stochastic gradient descent (SGD) with the descent step size $\eta$ to the substituted objective yields the self-supervision step

$$
\begin{aligned}
\mathbf{z}^n &= \mathbf{a}(\mathbf{f}(\mathbf{u}^n|\boldsymbol{\Theta}^n)) && \text{(denoising)}\\
\boldsymbol{\Theta}^{n+\frac{1}{2}} &= \boldsymbol{\Theta}^n - \frac{\eta\lambda}{2}\frac{\partial\|\mathbf{z}^n - \mathbf{f}(\mathbf{u}^n|\boldsymbol{\Theta})\|_2^2}{\partial\boldsymbol{\Theta}}(\boldsymbol{\Theta}^n)
\end{aligned} \quad (5)
$$

followed by the usual supervision step

$$\boldsymbol{\Theta}^{n+1} = \boldsymbol{\Theta}^{n+\frac{1}{2}} - \eta\frac{\partial D(\mathbf{y}^n, \mathbf{f}(\mathbf{x}^n|\boldsymbol{\Theta}))}{\partial\boldsymbol{\Theta}}(\boldsymbol{\Theta}^n), \quad (6)$$

in which $(\mathbf{x}^n, \mathbf{y}^n)$ is an image–label pair and $\mathbf{u}^n$ an unlabeled image (for clarity, we assume that the $n$th images and label are picked at the $n$th step). Updates (5) denoise $\mathbf{f}(\mathbf{u}^n|\boldsymbol{\Theta}^n)$ to produce a soft target $\mathbf{z}^n$ and update $\boldsymbol{\Theta}^n$ to reduce the $\ell^2$ loss between the reconstruction $\mathbf{f}(\mathbf{u}^n|\boldsymbol{\Theta}^n)$ and soft target $\mathbf{z}^n$. This is followed by another update to reduce the supervised loss between $\mathbf{y}^n$ and $\mathbf{f}(\mathbf{x}^n|\boldsymbol{\Theta}^n)$. Alternatively, the two updates can be performed in a single backward pass (Algorithm 1).

RED-induced update steps (5) force soft target $\mathbf{z}^n$ to be a function of only the current reconstruction and additionally restricts the loss to an $\ell^2$ one. However, it is known [23] that for stochastic optimization, averaging predictions across the training steps is crucial to reducing stochastic noise, hinting that the general form of the denoising step in (5) may not be optimal. Indeed, we find that iterating update steps (5) leads to divergent training outcomes in practice. Besides, several loss functions have previously been proposed specifically to solve segmentation-type imaging problems [15] but original RED would implicitly use the MSE loss on denoised targets.

## 3.3 Supervision by Denoising (SUD)

Supervision by denoising (SUD) seeks to combine RED and stochastic averaging used in semi-supervised learning (SSL) techniques [34], [35]. While SUD works with any recent SSL backbone, we assume that temporal ensembling [23] is used to facilitate analysis. This leads to a particularly elegant SUD self-supervision step of the form

$$
\begin{aligned}
\mathbf{z}^n &= \operatorname{prox}_{\varsigma R}(\alpha\mathbf{f}(\mathbf{u}^n|\boldsymbol{\Theta}^n) + (1-\alpha)\mathbf{z}^n)\\
\boldsymbol{\Theta}^{n+\frac{1}{2}} &= \boldsymbol{\Theta}^n - \eta\lambda\frac{\partial D(\mathbf{z}^n, \mathbf{f}(\mathbf{u}^n|\boldsymbol{\Theta}))}{\partial\boldsymbol{\Theta}}(\boldsymbol{\Theta}^n),
\end{aligned} \quad (7)
$$

in which $\operatorname{prox}_{\varsigma R}(\mathbf{f}) = \operatorname{argmin}_{\mathbf{t}} \varsigma R(\mathbf{t}) + (1/2)\|\mathbf{t} - \mathbf{f}\|_2^2$ denotes the "proximal map" associated with $\varsigma R$, $D$ is a loss function (cross entropy, Dice, etc.), $\mathbf{z}^n$ (rhs) is the last value of the soft target, and $0 \leq \alpha \leq 1$ specifies a combination weight for the current reconstruction and $\mathbf{z}^n$. In Appendix A, we derive the update steps (7) as a stochastic variant of proximal forward backward splitting [65] of a training objective that seeks the optimal soft targets and optimal network weights. Note that training of a fully convolutional network is a non-convex problem and convergence is only to a locally optimum.

Although update steps (7) can readily be justified from a proximal optimization perspective, the presence of the prox operator also necessitates solving an optimization problem at every training step, which can be costly. Since $\mathbf{a}$ is usually arbitrary (its objective is not to model physical phenomena but simply to denoise), we approximate the proximal step in (7) as a direct denoising step with $\varsigma = \alpha\beta$ (Appendix A):

$$\mathbf{z}^n = \alpha(\beta\mathbf{a}(\mathbf{f}^n) + (1-\beta)\mathbf{f}^n) + (1-\alpha)\mathbf{z}^n \quad (8)$$

in which parameter $0 \leq \beta \leq 1$ specifies the tradeoff between the smoothness of $\mathbf{z}^n$ and its fidelity to $\mathbf{f}^n = \mathbf{f}(\mathbf{u}^n | \mathbf{\Theta}^n)$. With $\beta = 0$, denoising step (8) reduces (7) to a variant of temporal ensembling [23]. The RED denoising step of (5) can be seen as a limiting case of (8) where $\alpha = \beta = 1$.

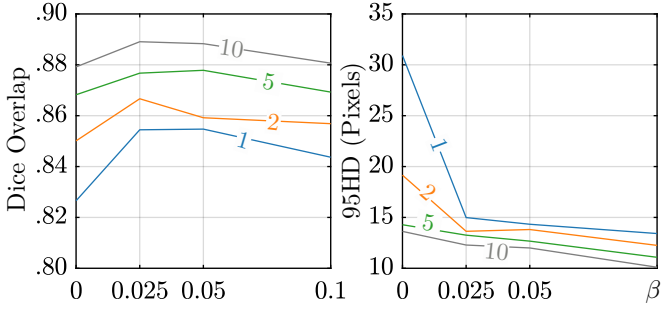

Fig. 5. Effect of varying $\beta$ on validation accuracy (Dice and 95HD) for the cortical parcellation task. Numbers 1, 2, 5 and 10 on lines denote the number of labeled training images used. Dice quickly improves with increasing $\beta$ and gradually worsens past $\beta = 0.05$. 95HD improves very quickly initially then gradually plateaus past $\beta = 0.05$. Plotted for $\lambda = 4$.

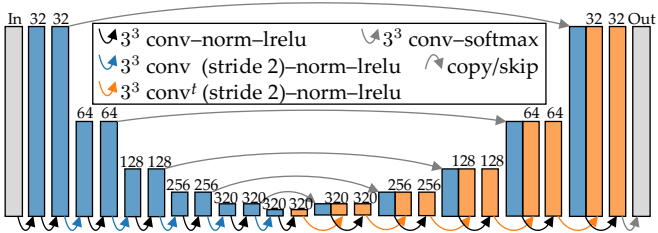

Fig. 6. Network structure used. Both our reconstruction and denoising networks have 5–6 levels of convolutions, instance normalization and leaky relu. We double the number of features at each new level for a maximum of 320 features. For denoisers, we remove skip connections and use max (un)pooling instead of strided convolutions.

Implemented naively, the simplified proximal step of (8) requires storage of the last value of the pseudo-target $\mathbf{z}^n$ for each unlabeled example, which can become very costly with a large unlabeled dataset. This storage requirement is easily addressed using an alternate approach, where an averaged network is used to generate pseudo-targets. In this case,

$$\mathbf{z}^n = \alpha\beta\mathbf{a}(\mathbf{f}^n) + (1-\alpha\beta)\mathbf{f}(\mathbf{u}^n | \overline{\alpha}\mathbf{\Theta}^n + (1-\overline{\alpha})\mathbf{\Theta}^{n-1}) \quad (9)$$

in which $\overline{\alpha} = (\alpha - \alpha\beta)/(1 - \alpha\beta)$. With $\beta = 0$, denoising step (9) reduces to the weight averaging of the mean teacher [35] model. We find that either denoising step can be used in the context of SUD with no noticeable difference in the accuracy of the trained reconstruction network. We now analyze the spatial and the temporal filtering aspects of SUD, governed respectively by parameters $\beta$ and $\alpha$.

### 3.3.1 Spatial Denoising Aspects

The spatial filter parameter $\beta$ bridges the gap between direct and optimization-based (proximal) denoising methods. Let us assume $\alpha = 1$ and write $\mathbf{z}_d^n = \mathbf{a}(\mathbf{f}^n) = (J\mathbf{a}(\mathbf{f}^n))\mathbf{f}^n = \mathbf{A}\mathbf{f}^n$ to express direct denoising of an iterate $\mathbf{f}^n$, invoking the "local homogeneity" [28] of $\mathbf{a}$ (we assume $\mathbf{a}$ is differentiable solely for analysis). Spectral decomposition $\mathbf{A} = \mathbf{U}(\mathbf{I} - \mathbf{\Lambda})\mathbf{U}^*$ tells us that $\mathbf{A}$ has spectral filter factors [47]

$$1 \geq 1 - \lambda_1 \geq 1 - \lambda_2 \geq \cdots \geq 1 - \lambda_N \geq 0, \quad (10)$$

in which $\lambda_n$ denotes the $n$th diagonal element of $\mathbf{\Lambda}$, the first inequality comes from the assumed strong passivity of $\mathbf{a}$ [28] and the last one comes from its positive definiteness.

For optimization-based denoising, consider evaluation of the mapping $\text{prox}_{\beta R}$ (7) on $\mathbf{f}^n$. If we linearize the denoiser in the $R$ term as $\mathbf{a}(\mathbf{u}) \approx (J\mathbf{a}(\mathbf{f}^n))\mathbf{u} = \mathbf{A}\mathbf{u}$, the proximal mapping can be expressed as $\mathbf{z}_p^n = (\mathbf{I} + \beta(\mathbf{I} - \mathbf{A}))^{-1}\mathbf{f}^n = \mathbf{H}\mathbf{f}^n$. Spectral decomposition can be used similarly to reveal that $\mathbf{H}$ shares the same spectral filter basis $\mathbf{U}$ as $\mathbf{A}$ but has filter factors

$$1 \geq \frac{1}{1 + \beta\lambda_1} \geq \cdots \geq \frac{1}{1 + \beta\lambda_N} \geq (1+\beta)^{-1} \quad (11)$$

so the denoisers $\mathbf{A}$ and $\mathbf{H}$ differ only in their filter factors. In particular, (11) shows $\mathbf{z}_p^n$ is a linear combination of denoised input and the unfiltered input with weights $\beta(1+\beta)^{-1}$ and $(1+\beta)^{-1}$ respectively. Fig. 4 (left) shows this decomposition in terms of spectral filter factors (11).

SUD essentially re-parameterizes the weights $\beta(1+\beta)^{-1}$ and $(1+\beta)^{-1}$ as $\beta$ and $(1-\beta)$. Weights $\beta, 1 - \beta$ are used to combine the unfiltered $\mathbf{f}^n$ and denoised $\mathbf{a}(\mathbf{f}^n)$ components to produce a soft target $\mathbf{z}_d^n$ that approximates the proximal one $\mathbf{z}_p^n$. This approximation obviates the need to solve proximal optimization problems at train time and has more flexibility than that used by RED [27], which implicitly constrains the soft target to $\mathbf{z}_d^n = \mathbf{a}(\mathbf{f}^n)$. Fig. 5 plots reconstruction accuracy under SUD for different values of $\beta$, showing that optimally combining $\mathbf{a}(\mathbf{f}^n)$ and $\mathbf{f}^n$ is beneficial for supervision.

### 3.3.2 Temporal Denoising Aspects

At train time, random noise in the predicted reconstruction due to augmentation and dropout may cause soft targets to change from one training step to the next, creating a moving target phenomenon. Therefore, denoising soft targets across iterations helps average out stochastic prediction noise and improves training convergence [23]. The filter parameter $\alpha$ can be seen equivalently as defining a first order IIR (infinite impulse response) filter with the temporal impulse response

$$(\alpha, \alpha(1-\alpha)^1, \ldots, \alpha(1-\alpha)^n, \ldots), \quad (12)$$

also called an exponential moving average filter [35]. Note that an IIR filter of order $k$ necessitates additionally storing the last $k$ soft targets for each training image. Implementing a higher order temporal IIR filter is thus more costly and yet provide no clear averaging benefits over a first order one. In the case of a first order filter, temporal filtering can be recast equivalently as a proximal gradient descent step where the descent step size is $\tilde{\alpha} = \alpha/\lambda$; see Appendix A.

In practice, we graduate the filter parameter $\alpha$ from 1 to 0 across the course of the training, enabling the soft targets to adapt to the rapidly evolving nature of $\mathbf{f}(\mathbf{x}|\mathbf{\Theta})$ at the start of training and converging to fixed (non-moving) targets at the end of the training. Fig. 4 (right) plots the impulse response of the time-varying denoising filter, depicting the influence of current soft targets on future iterations. Varying $\alpha$ can be motivated also from an optimization perspective [65], where decaying step sizes are commonly used for convergence. We find that setting $\alpha = 0.9$ or $0.99$ as in [35] slightly decreases performance—unlike classification networks, which output a pooled prediction across the entire image, reconstruction networks produce a prediction at each individual pixel with a larger temporal prediction variance as a result. Decreasing the value of $\alpha$ agressively towards zero ensures that the soft targets converge. We vary $\lambda$ in the opposite manner, from 0 to $\lambda_{\max}$, bringing self-supervision gradually into effect. This is similar to weight ramping in [23], which uses a Gaussian

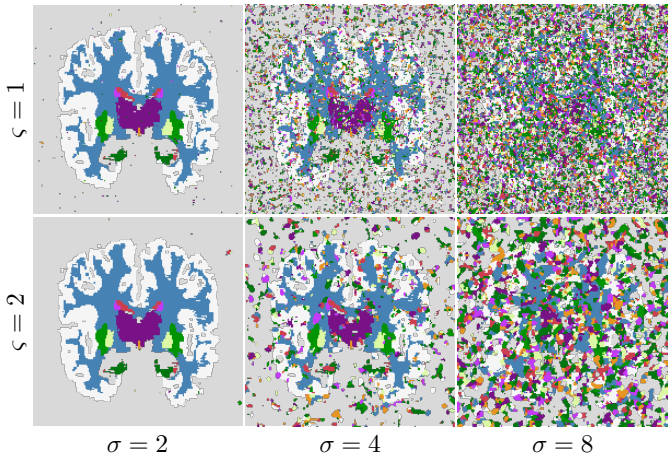

Fig. 7. Reconstruction example corrupted by Gaussian noise of varying standard deviation $\sigma$ and scale $\varsigma$ (argmax of classes shown). Increasing $\sigma$ simply amplifies the magnitude of the noise injected. Increasing $\varsigma$, on the other hand, magnifies the spatial scale of the noise. When both are increased, the denoiser is forced to learn to inpaint rather than denoise.

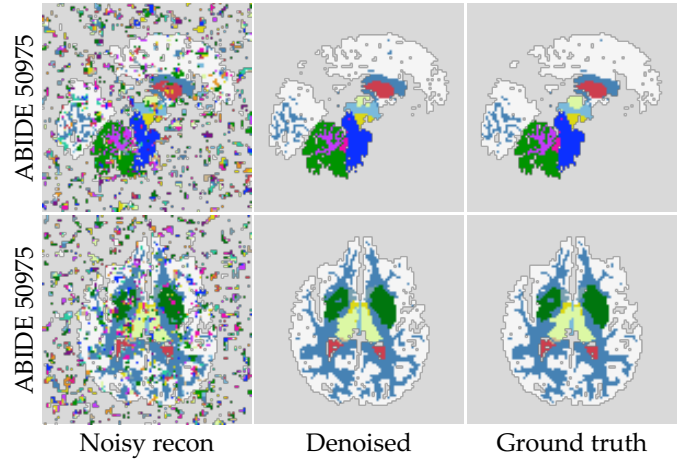

Fig. 8. Examples of noisy reconstructions (left column), denoised by our auto-encoder denoisers (center column), together with the ground truth (right column). Even in the absence of skip connections, auto-encoder-based denoisers provide good localization of reconstruction boundaries comparable to their U-Net-based counterparts.

(as opposed to our linear) schedule for varying $\lambda$.

## 3.4 Network Structures and Training

A popular choice of reconstructor is U-Net [5], which is also used in this work. Following nnU-Net [36], we configure our U-Net with five levels of convolutions and leaky ReLU, and double the number of feature channels at each new level up to a maximum of 320 features; see Fig. 6 for illustration.

The main motivation behind SUD, as well as RED [27], is to leverage recent development in learned denoising filters to solve imaging and vision problems. In this work, we train a secondary label-denoising network $\mathbf{a}(\cdot\,|\boldsymbol{\Sigma})$ for supervision of the main reconstruction network $\mathbf{f}(\cdot\,|\boldsymbol{\Theta})$. From a Bayesian perspective, denoiser network $\mathbf{a}$ defines a log-prior on $\boldsymbol{\Theta}$ via the regularity term $R(\mathbf{f}(\mathbf{x}|\boldsymbol{\Theta}))$ while reconstructor $\mathbf{f}$, the log-likelihood $D(\mathbf{y}, \mathbf{f}(\mathbf{x}|\boldsymbol{\Theta}))$ of $\boldsymbol{\Theta}$ given $(\mathbf{x}, \mathbf{y})$. The denoiser plays an important role when learning network parameters $\boldsymbol{\Theta}$ by solely optimizing $D(\mathbf{y}, \mathbf{f}(\mathbf{x}|\boldsymbol{\Theta}))$ may be challenging. In many problems, a prior over the reconstructions (denoiser) can be easier to learn than directly optimizing the likelihood term by training the reconstruction network. This is often the case if the reconstructions live in the space of natural images, or medical image segmentations, for which previous reference reconstructions from other datasets exist to train a denoiser.

As for the choice of denoising network architecture, both convolutional (denoising) auto-encoder [51] and U-Net ones [66] have previously been used, with the main architectural difference between the two being the shortcut connections in the latter. The shortcut connections allow the U-Net to learn subtractively and predict the noise or "residuals" [67] in the reconstruction while auto-encoders must learn to predict the true reconstruction directly. Therefore, a U-Net denoiser can no longer be trained on reconstructions which are corrupted by artificial Gaussian noise—the noise must be synthesized to emulate artifacts in the output of the reconstructor under training, which can cause further training challenges for the denoiser. While auto-encoders lack skip connections, we do not find that this leads to worse localization of boundaries in the denoised reconstruction. Auto-encoders tend to perform empirically better than the U-Net ones in the reconstruction experiments. The observation in [68], [53] also highlights the inability of U-Net to inpaint images if not trained explicitly to impute missing regions, further suggesting that learning the latent space of reconstruction using auto-encoders leads to better regularization and thus semi-supervision. For this reason, we use auto-encoder denoisers exclusively in all our experiments. Our denoisers share the same structure as our U-Net reconstructor with the skip connections removed and the max pooling replaced by strided convolutions.

To generate training data for denoising, we log-transform the given one-hot label (segmentation) maps, clip them to a $[-3, 0]$ range ($-3$ and $0$ corresponding to the probabilities of $0$ and $1$ respectively), corrupt them using additive Gaussian noise of zero mean and random standard deviation $\sigma$, then finally take their softmax. To generate noise across multiple scales, we randomly downsample produced Gaussian noise map by $\varsigma$ and upsample it to the original scale. Fig. 7 shows examples of randomly generated noisy training images with different values of the variance–scale pairs $(\sigma, \varsigma)$. When both $\sigma$ and $\varsigma$ are large, structure appears in the noise pattern and the denoising task becomes similar to that of inpainting.

In many problem domains, label maps needed to train a denoiser can be obtained from computer games [69], [70] or even consist of arbitrary shapes [71] to potentially generate unlimited training data. Many times, however, the denoiser must itself be learned on limited label map data to bootstrap semi-supervision. In Table B4, we list the mean Dice overlap and 95HD attained by the denoisers with different numbers of training label maps, with augmentation hyperparameter ranges $\varsigma, \sigma \in [0,8]$. The table shows that very few examples are needed to train a denoiser. In Fig. 8, we show the noisy reconstruction examples used for validation, along with the corresponding output from the auto-encoder denoisers. We pretrain our denoisers on reconstruction examples and use them with weights fixed when training the reconstructor.

## 4 EXPERIMENTS

We validate SUD by applying it on two biomedical imaging problems: anatomical brain reconstruction (3D) and cortical parcellation (2D). See Fig. 9 for an illustration of the cortical parcellation task. We use the nnU-Net [36] variant of U-Net and image augmentation strategies (see Appendix B) for a fair evaluation against the supervised U-Net baselines. Our "post-denoise" method implements the method of [53] and denoises baseline reconstructions using SUD's denoiser. The "ensembling" method is similar to [33], [34] with a temporal ensembling backbone for averaging. We use hyperparameter settings $\beta = 0.2/\lambda_{max}$ and $\lambda_{max} = 4$ in all experiments and $\alpha$ is varied linearly from 0 to 1 across training steps. Methods such as MixUp [72] and CutMix [73] are intended for image classification tasks and are not compared with in this work.

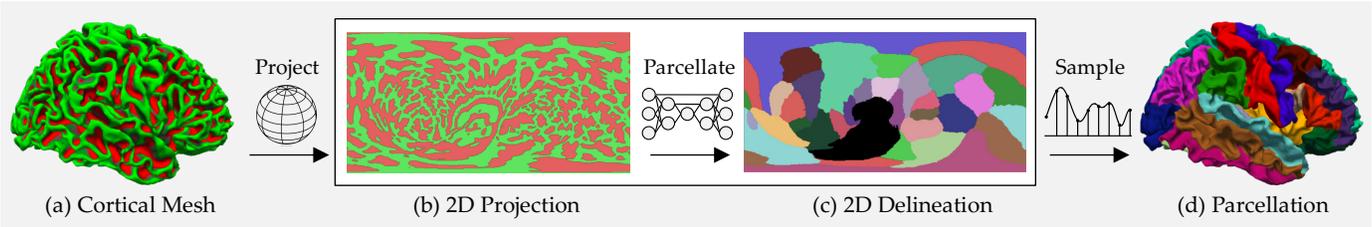

Fig. 9. Cortical parcellation pipeline. Given a cortical mesh model (a), we project it to a 2D image (b) and parcellate it to produce a parcellation map (c). The parcellation map is then sampled onto the mesh model to produce the final parcellation. Since the projection and the sampling are fixed operations, we treat parcellation as a 2D visual reconstruction problem (highlighted in white).

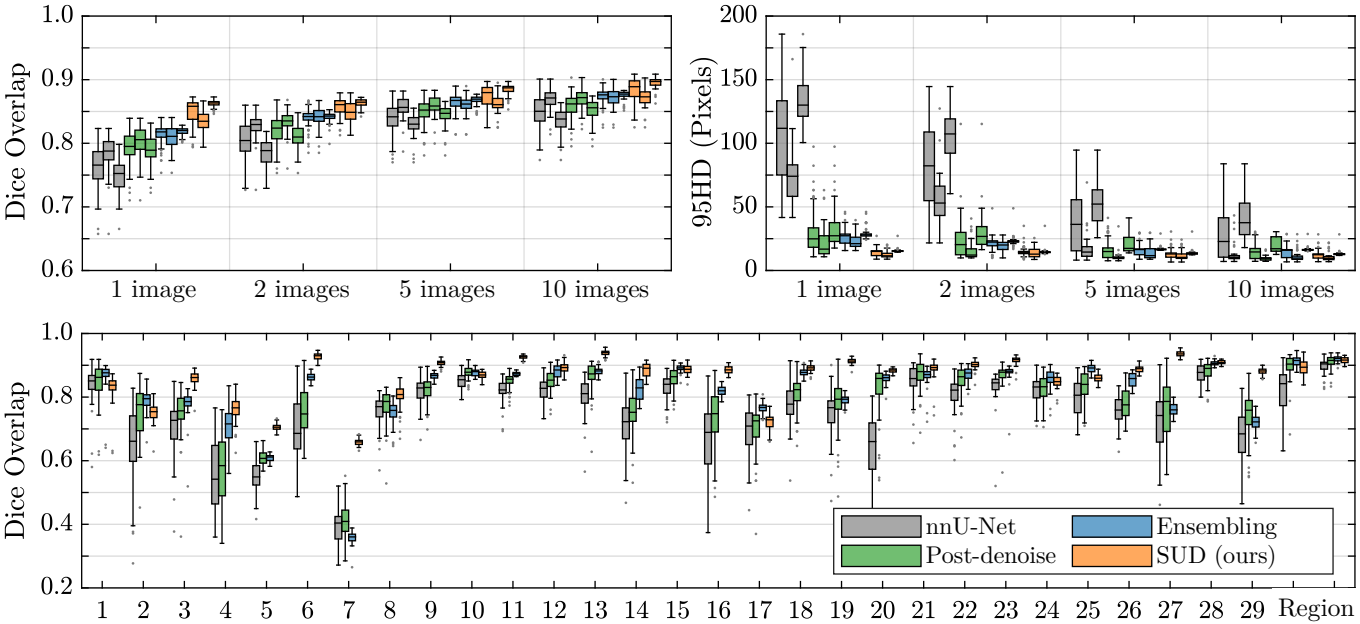

Fig. 10. Dice and 95HD statistics (test) for the cortical parcellation task. The top row plots the statistics for different training regimes (nnU-Net, with additional denoising, temporal ensembling, and SUD), with each group of three box plots representing the statistics on the MindBoggle + HCP (90 test subjects), MindBoggle (40 test subjects) and HCP (50 test subjects). The bottom row plots the Dice overlap of the individual cortical regions for the HCP dataset when trained using one labeled image. See Fig. 11 (top left) for a diagram of the cortical regions.

In the anatomical brain reconstruction problem, we show that SUD improves the reconstructor's ability to generalize to images produced under vastly different image acquisition hardware. On the other hand, in the 2D cortical parcellation task, we demonstrate that SUD is extremely useful when the likelihood term (1) is unreliable due to lack of image texture.

## 4.1 Brain Cortical Parcellation (2D)

Segmenting the human cerebral cortex into cortical areas of interest (known as parcellation, Fig. 9) is a problem of great importance in neuroscience [74]. However, the challenging nature of labeling white matter and pial surfaces leads to the unavailability of labeled cortical meshes for training. In this task, image texture provides less useful reconstruction cues (Fig. 9, b) than in the 3D brain reconstruction task so a large amount of labeled training data may be required to achieve generalization. We focus on the reconstruction portion of the parcellation pipeline (highlighted in white, Fig. 9) since the projection and the sampling steps are relatively simple pre- and post-processing operations and do not require learning.

In this experiment, we leverage SUD with a large dataset of unlabeled human cortical models to learn a more robust parcellation model. For labeled data, we use the Mindboggle dataset [75] containing 101 subjects, annotated in 31 cortical regions. We also use the Human Connectome Project (HCP) [76] image for training via SUD. We split the HCP data into train and testing such that there is no overlap in the subjects between training images for the reconstructor and training labels for the denoiser. We pick the network of the last epoch for all training methods except the fully supervised nnU-Net baseline, where the network of the best epoch is chosen. See Fig. 3 for issues with using the last epoch nnU-Net model.

We use the (sulcal map, white matter curvature, inflated surface curvature) data, generated using FreeSurfer [77], as input, apply a 98% Winsoring to the white matter curvature to remove outliers and normalize each image channel to unit standard deviation and zero mean. We augment image data using 2D random elastic deformation. The deformation field is generated by smoothly interpolating displacement vectors defined on a grid with $16 \times 16$ cells. We train all models for a maximum of 200,000 iterations and a batch size of 1. We use the Adam optimizer with the learning rate of $0.0001$, weight decay of $0$ and epsilon of $10^{-8}$. Channel dropout with keep probability $0.95$ is used to reduce overfitting that manifests especially egregiously in this cortical parcellation problem.

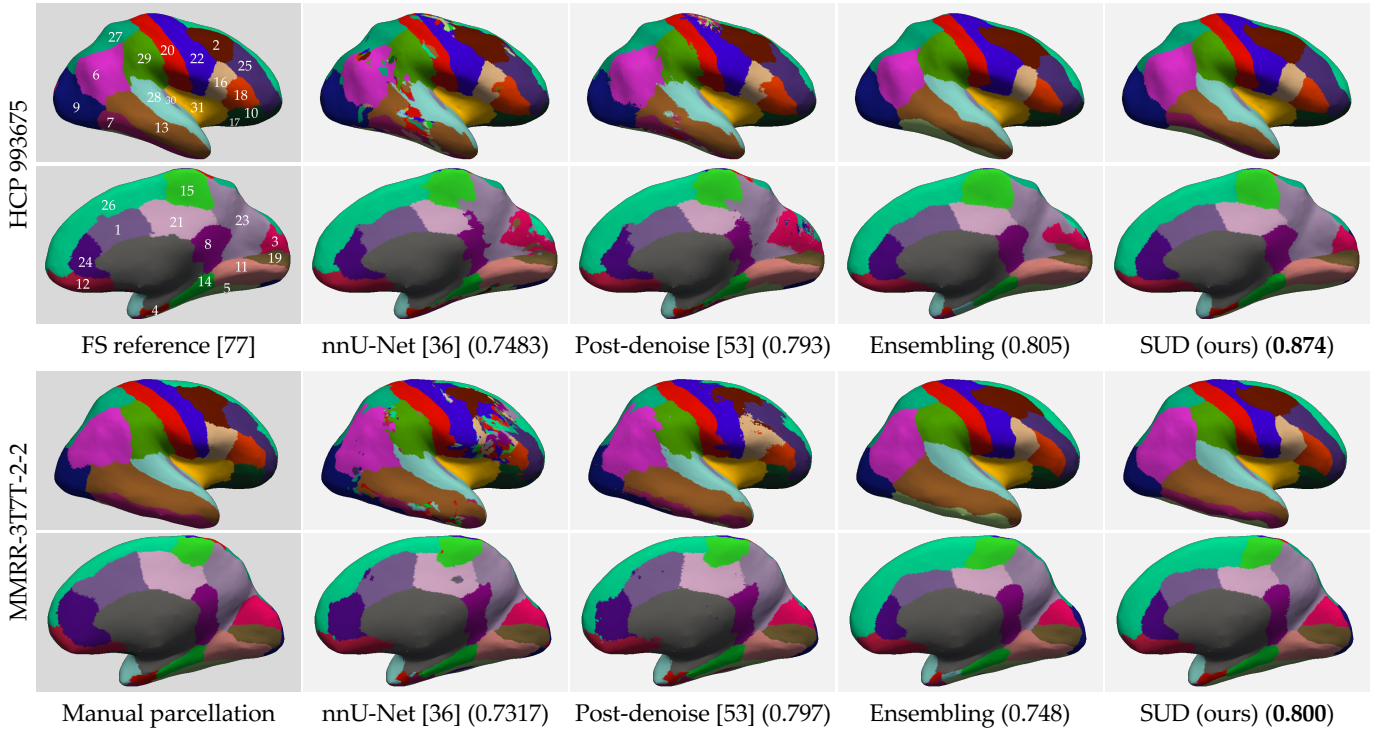

Fig. 11. Cortical parcellations (test) produced under different training regimes, shown together with FreeSurfer reference ones (mean Dice overlap in parentheses are with respect to the reference). All training regimes use only one labeled image and share the same underlying nn-UNet implementation. Post-denoise and SUD are based on the same denoiser. SUD produces the most visually accurate parcellations as well as the highest mean Dice scores. See Fig. 10 for the overall statistics. Rendered using FreeView. Figure best viewed online.

Fig. 10 plots the mean Dice and Hausdorff distance of the reconstructions predicted by U-Net under different training regimes, supervised baseline, stochastic ensembling, as well as SUD, across 1, 2, 5 and 10 labeled images. When 1 labeled image is used for training, SUD U-Net improves mean Dice and Hausdorff Distances over stochastic ensembling by 0.04 and 18 pixels (~54%), respectively. Most of the improvement comes from the HCP dataset, for which no labeled training images are available. Region-wise, the largest improvements are from 7 (inferior temporal lobe), 27 (superior parietal) and 29 (supra-marginal) with Dice gains ranging $0.2$–$0.3$. Fig. 11 shows parcellations predicted under each of the supervision regimes. The baseline U-Net predictions exhibit an excessive amount of reconstruction noise but are somewhat improved by post-denoising. Stochastic ensembling misses all inferior temporal lobes in its parcellations while SUD predicts them accurately without any obvious reconstruction noise.

### 4.2 Anatomical Brain Reconstruction (3D)

Anatomical reconstruction of brain MR scans is another task of importance in neuroimaging as it enables volumetric and morphological analyses of the brain in healthy and diseased populations [78]. However, the 3D volumetric nature of MR scans renders manual delineation of the anatomical regions challenging, necessitating many days of labeling efforts per image. Moreover, the differences in MRI scanners and their scanning protocols can hamper the reuse of a reconstruction network trained on a particular dataset or image contrast. In this case, we can train a denoiser on previous reconstruction examples and semi-supervise the reconstructor via SUD.

In this experiment, we apply SUD to the training of a 3D brain reconstruction model that is more robust to variations in the MRI setting. We use only one labeled image similarly to the one-shot reconstruction approach of Zhao *et al.* [13] but rather than improve generalization by learning the optimal data augmentation as in [13], we achieve this by training on unlabeled scans—ABIDE-1, -2, ADHD200, GSP, MCIC, UK BIO, COBRE and PPMI—as used in [79]. Refer to Table B2 in Appendix B for the exact breakdown. All images have been rotated into a common orientation. We train the denoiser on 1000 label maps generated using FreeSurfer [77]. Similar to the cortical parcellation experiment, the 1000 label maps do not have subject overlaps with any of the images used, and the validation data is used only for assessing divergence or potential overfitting of the supervised nnU-Net baseline. We optimize the augmentation (shown in Table B1) and train for 200,000 iterations using the Dice loss under all regimes. We use identical optimization settings as in in cortical parcellation.

Fig. 12 plots the distribution of mean Dice and Hausdorff distances of U-Net reconstructions produced under different supervision regimes, conditioned on subject sets, as well as on brain structures. All supervision regimes produce similar accuracy results (95HD in particular) on Buckner39 since the labeled training image is sourced from this set. The accuracy gap for other subject sets, especially ABIDE2 and ADHD, is considerably larger, where the improvement in median Dice due to SUD is ~0.05. To see the improvement in Dice and 95 Hausdorff distance at the individual subject level, we plot in Fig. 13, histograms of Dice and 95HD change across the 500 test subjects between SUD and ensembling methods, as well as SUD and post-denoised. The mean Dice improvement is strictly positive with peak improvement of 0.15 while losses in the individual mean 95HD are observed sometimes. This is somewhat expected since our training objective explicitly optimizes Dice. The 95HD improvement is still a statistically significant one ($p = 10^{-37}$, Wilcoxon's signed-rank test).

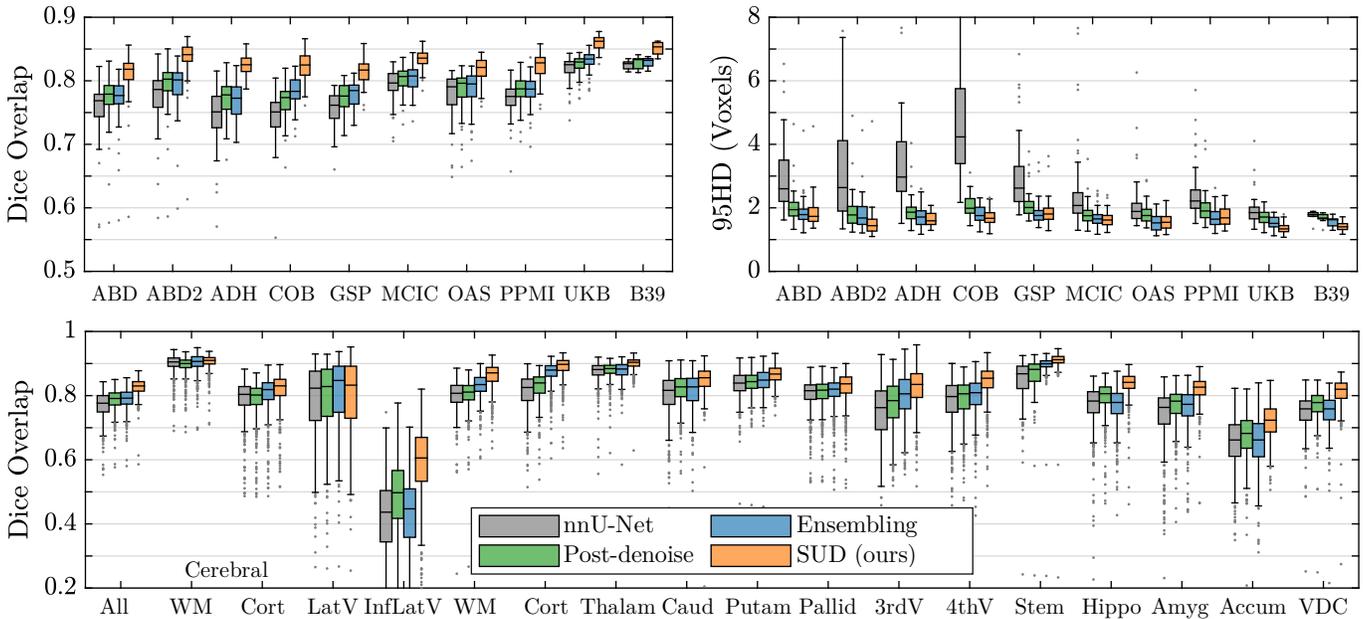

Fig. 12. Dice and 95HD statistics (test) for the brain reconstruction task, conditioned on individual datasets (top) and brain structures (bottom). We plot the two metrics for the supervised nnU-Net baseline (trained on one labeled Buckner39 image), with additional post-denoising, stochastic ensembling, and SUD (ours). See text for details on the individual datasets.

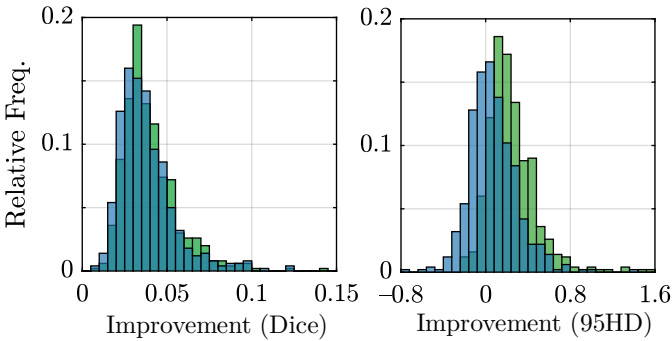

Fig. 13. Dice and 95HD improvements (test) due to SUD, relative to the "stochastic ensembling" (blue) and "post-denoise" (green) methods. The same nnU-Net model is used in all three cases. Plotted across 500 test images. The Dice improvement is strictly positive.

Fig. 14 visualizes the 3D anatomical brain reconstructions produced by U-Net under the different training regimes. We include reconstructions produced using FreeSurfer [77]. We choose subjects that best demonstrate improvements due to SUD with a Dice gain in the range $0.06 - 0.08$. A more typical Dice gain is around $0.05$, as is seen in Fig. 13 (left). Smaller structures (hippocampus, amygdala, etc.) are visually better reconstructed, as is the lateral ventricle, with fewer holes in the SUD reconstructions. Dice improvements in the inferior lateral ventricle—as indicated by Fig. 12—are not so clearly observed in the visualization. We note that the Dice scores are computed against the FS reference reconstructions and should be interpreted with caution—only the human eye is able to gauge the quality across the various reconstructions.

## 5 DISCUSSION

Our experimental results can be summarized as follows:

- In supervision regimes with very few labeled images (one in the case of anatomical brain reconstruction and cortical parcellation), SUD improves the reconstruction Dice score by 5 points on average over stochastic ensembling and by 10 points against the fully supervised baseline. The 95HD is reduced by 50–75% over stochastic ensembling.
- For 3D brain reconstruction, the Dice improvement due to SUD over stochastic averaging is attributed to subsets of images for which no labels are available (e.g., ABIDE2, 5 points). Improvement on Buckner39, where the labeled training data was sourced from, is less (1 point).
- Post-denoising a predicted reconstruction does not yield the same Dice improvement as training the reconstruction network under SUD. This is similar to classical inverse problems (e.g. image deblurring), where a naive solution cannot be denoised post-hoc due to inversion noise.

### 5.1 Extending SUD

Extensions to temporal ensembles and mean teacher models [60], [61] involve training multiple teacher networks where the weights are initialized differently. The output from the teachers is then used as pseudo-targets for the other during training. Training multiple networks leads to an increase in the training time and resources, but these ideas may still be useful for providing better stochastic averaging in SUD and potentially futher improving semi-supervision. Rather than exhaustively investigate the optimal number of networks to train for stochastic averaging, here we focus instead on the usefulness of spatial denoising as a supervision mechanism.

Advanced image augmentation strategies such as MixUp [72], and CutMix [73] combine two images by addition, and stitching, respectively. These provide an easy way to further improve the diversity of labeled data for visual classification (i.e. non-reconstruction) problems. Whereas simple intensity and geometry-based image augmentation strategies suffice in our two reconstruction tasks, aggressive augmentation of datasets can be useful especially with too few labeled images to bootstrap semi-supervised learning. However, adapting MixUp and CutMix to an image reconstruction task is not so straight-forward and we defer their discussion to a sequel.

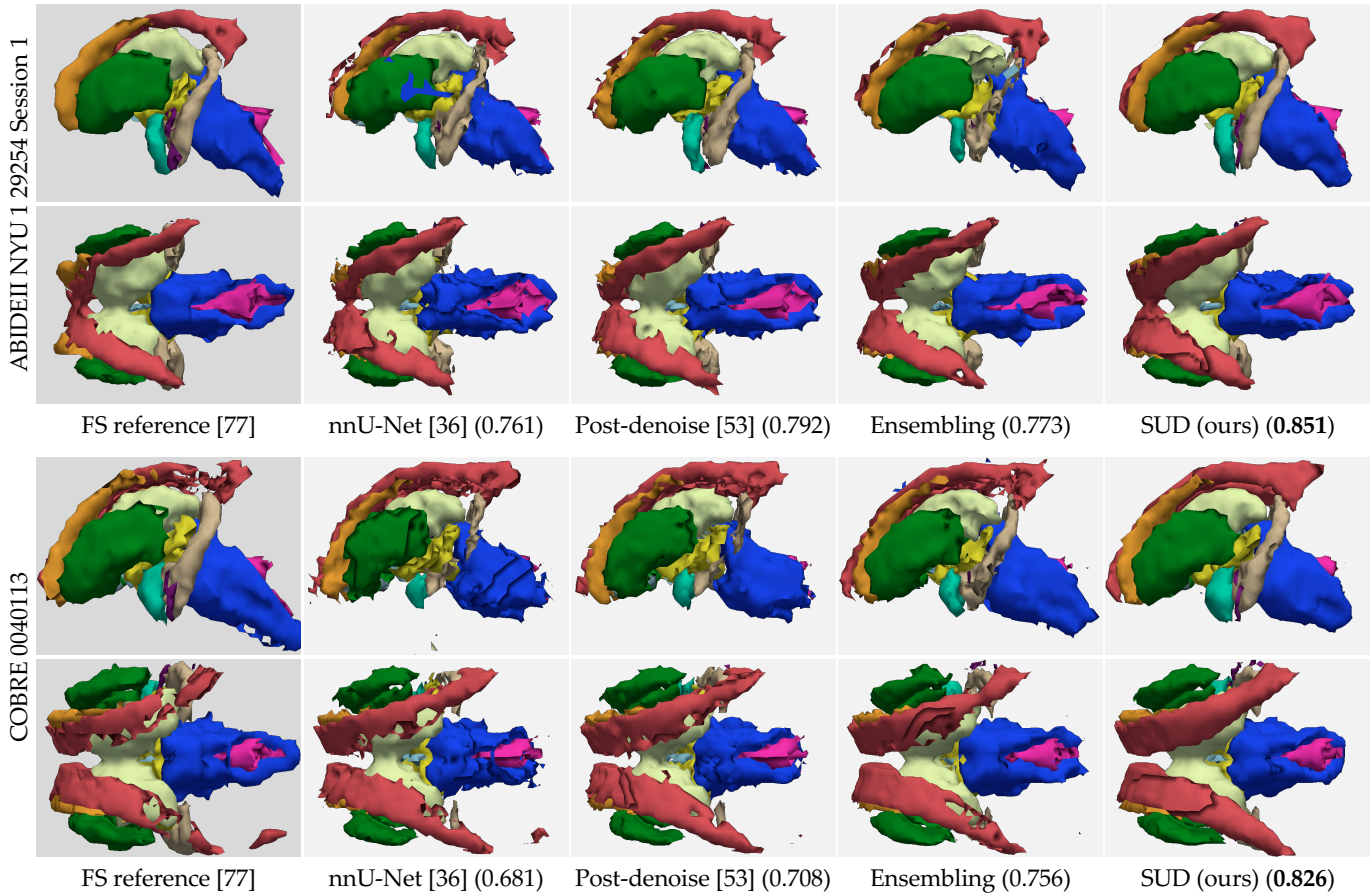

Fig. 14. Brain structure reconstructions (test) produced under different training regimes, shown together with reference FreeSurfer ones (mean Dice overlap in parentheses with respect to the reference). Cortex and white matter not shown for clarity. All training regimes use only one labeled image and share the same underlying nn-UNet implementation. Post-denoise and SUD use the same auto-encoder denoiser. SUD produces the most visually accurate reconstructions as well as the highest mean Dice scores. See Fig. 11 for the overall statistics. Rendered using FreeView.

## 5.2   Possible Limitations

A core premise of SUD is that learning reconstruction priors (that is, training a denoiser) is, in a sense, easy compared to optimizing the likelihood (training the reconstructor). Such an assumption is reasonable and implied in many imaging and visual reconstruction tasks, where a denoiser is readily trained on plenty of existing reconstructions. In tasks where few reconstruction examples exist to begin with, learning a prior over reconstructions can be difficult owing to a general lack of knowledge that can be "distilled" into a prior. Our assumption of "sufficiently many and varied reconstruction examples" is far from being unique to our work, and serves as the basis in computational photography [1], [2], in which pretrained natural image denoisers are used and biomedical imaging [79], [71], [53] where many reconstruction examples facilitate denoising and generation of reconstructions.

## 6   CONCLUSION

In learning-based image reconstruction, semi-supervision or improving the generalization of a reconstruction network by additionally learning from unlabeled images is a problem of theoretical and practical interest. In this work, we proposed supervision by denoising (SUD), enabling semi-supervision using the reconstruction network's own (denoised) output as pseudo-targets. SUD is closely related to regularization by denoising (RED) used for solving inverse problems, which highlights again that many problems in imaging and vision including supervision ones, can be ensconced in a denoising framework. Under the SUD framework, such techniques as temporal ensembling and mean teacher can be reinterpreted as a form of SUD that does not involve a spatial denoiser.

Given the rather generic nature of SUD, we believe SUD and potentially its mean teacher variant can serve as a useful tool for semi-supervised reconstruction especially if used in tandem with more advanced data augmentation techniques and task-specific strategies training denoisers. It is our hope that our work will be broadly applicable to all reconstruction problems faced by imaging and vision researchers, not just those specifically mentioned in this work.

## ACKNOWLEDGMENTS

This work is primarily supported the NIH BRAIN Initiative (RF1MH123195). Additional support is provided by the NIH (U01MH117023, R01EB023281, R01EB006758, R01EB019956, R01AG064027, RF1MH121885, R01NS083534, R01NS105820, P41EB015902), ARUK (IRG2019A-003), ERC (Starting Grant 677697). EF was funded by the Fulbright-CONICET Visiting Researcher Program and he acknowledges the support of the UNL and ANPCYT. CAM was supported, in part, by AFOSR Young Investigator Program Award FA9550-22-1-0208.

**Sean I. Young** received his PhD degree in electrical engineering from the University of New South Wales, Sydney, Australia. He is currently an Instructor at the Martinos Center for Biomedical Imaging, Harvard Medical School, and a Research Affiliate in CSAIL, MIT. Previously, he was a postdoctoral researcher at Stanford University, Stanford, CA. In 2016, he was a visiting researcher at InterDigital Communications, San Diego, CA. He received the APRS/IAPR best paper award at DICTA 2018, together with David Taubman.

**Adrian V. Dalca** is an assistant professor at Harvard Medical School, Massachusetts General Hospital, and a Research Scientist in CSAIL, MIT. His research focuses on machine learning techniques and probabilistic models for medical image analysis. Driven by clinical questions, he develops core learning algorithms, as well as registration, segmentation and imputation methods aimed at clinical-sourced datasets and broadly applicable in image analysis.

**Enzo Ferrante** received his PhD degree in computer sciences from Université Paris-Saclay (Paris, France) and he worked as a postdoctoral researcher at the BioMedIA Lab, Imperial College London (London, UK). He is currently a CONICET faculty researcher and professor at Universidad Nacional del Litoral in Santa Fe, Argentina, where he leads the research line on machine learning for biomedical image analysis at the Research Institute for Signals, Systems and Computational Intelligence, sinc(i).

**Polina Golland** is a Professor of Electrical Engineering and Computer Science at MIT. Her research interests span computer vision and machine learning. Her current work focuses on developing statistical analysis methods for characterization of biological processes using images (from MRI to microscopy) as a source of information. She received BSc and Masters in Computer Science from Technion, Israel in 1993 and 1995, and a PhD in Electrical Engineering and Computer Science from MIT in 2001. She joined the faculty in 2003.

**Christopher A. Metzler** is an Assistant Professor of Computer Science at the University of Maryland, College Park, where he directs the UMD Intelligent Sensing Lab. He received his BS, MS, and PhD in electrical and computer engineering from Rice University in 2013, 2014, and 2019, respectively. He was a postdoc in the Stanford Computational Imaging Lab in 2019–2020. His work has received multiple best paper awards; he recently received an AFOSR Young Investigator Program Award; and he was an Intelligence Community Postdoctoral Research Fellow, an NSF Graduate Research Fellow, a DoD NDSEG Fellow, and a NASA Texas Space Grant Consortium Fellow.

**Bruce Fischl** is a Professor of Radiology at the Martinos Center for Biomedical Imaging and an affiliated researcher at MIT. His research involves the development of techniques for cortical surface modelling, thickness measurement, inter-subject registration, whole-brain segmentation and cross-scale imaging. The tools that have resulted from this research have been downloaded over 50,000 times and are in use in labs around the world.

**Juan Eugenio Iglesias** was born in Seville, Spain. He holds MSc degrees in Telecommunication and Electrical Engineering from the University of Seville and the Royal Institute of Technology (KTH, Stockholm, Sweden). He did a PhD in Biomedical Engineering at the University of California, Los Angeles (2007–2011), sponsored by a Fulbright grant. He was a postdoctoral researcher at the Martinos Center for Biomedical Imaging (2011–2014) and the Basque Center for Cognition, Brain, and Language (2014–2016), sponsored by a Marie Curie Fellowship. In 2016, he joined University College London as junior faculty, funded by Starting Grant of the European Research Council. In 2019, he rejoined the Martinos Center, where is now Associate Professor. He also holds honorary appointments at University College London and MIT.

# Appendix A    Mathematical Derivations

Here, we provide a brief derivation of the denoising step of (7) for two loss functions of interest: quadratic ($\ell^2$) loss and the Dice loss. The former facilitates analysis of the denoising step and comparison with direct denoising (8), whereas the latter a more general descent scheme for optimization-based denoising when a non-quadratic loss function is involved.

## A.1    Proximal Optimization and SUD

One alternative to objective (2) is to reformulate training as the joint optimization of the network parameters $\Theta$ and soft targets $\mathbf{z}^n$ for $n = 1, \ldots, U$. The corresponding objective is

$$F(\Theta, \mathbf{z}^{\{n\}}) = G(\Theta) + \frac{1}{U} \sum_{n=1}^{U} \lambda D(\mathbf{z}^n, \mathbf{f}(\mathbf{u}^n | \Theta)) + \tilde{\beta} R(\mathbf{z}^n) \quad (13)$$

noting that (13) is well-posed only if both terms involving $\mathbf{z}$ are present. The composite nature of objective (13) renders it amenable to optimization via a proximal forward-backward splitting method [65], which essentially alternates stochastic gradient descent and denoising steps.

In particular, the stochastic gradient descent step of (7) is used to minimize $F(\Theta, \mathbf{z}^{\{n\}})$ over $\Theta$ with $\mathbf{z}^{\{n\}}$ fixed. On the other hand, gradient descent to minimize $\lambda D(\mathbf{z}^n, \mathbf{f}(\mathbf{u}^n | \Theta))$ in the soft target $\mathbf{z}^n$ with weights $\Theta$ fixed can be written as

$$\mathbf{z}^n = \operatorname*{argmin}_{\mathbf{z} \in S} \lambda D(\mathbf{z}^n) + \langle \mathbf{z}, \lambda D'(\mathbf{z}^n) \rangle + \frac{1}{2\tilde{\alpha}} \|\mathbf{z} - \mathbf{z}^n\|_2^2 \quad (14)$$

[65], where $S$ is a direct sum of probability simplices across all pixel locations and $\tilde{\alpha}$ a descent step size.

Applying (14) to minimize $F(\Theta, \mathbf{z}^{\{n\}})$ over $\mathbf{z}^{\{n\}}$ amounts to finding $\operatorname{argmin} \tilde{\beta} R(\mathbf{z}) + \langle \mathbf{z}, \lambda D'(\mathbf{z}^n) \rangle + \frac{1}{2\tilde{\alpha}} \|\mathbf{z} - \mathbf{z}^n\|_2^2$. With the change of variables $\alpha = \tilde{\alpha}\lambda$ and $\beta = \tilde{\beta}/\lambda$, we can express this argmin as the proximal mapping in (7) when the loss is assumed quadratic: $D(\mathbf{z}, \mathbf{f}) = (1/2)\|\mathbf{z} - \mathbf{f}\|_2^2$. Direct denoising step (8) is derived by linearizing $R(\mathbf{z})$ at $\mathbf{f}^n$ and simplifying the resulting argmin expression to

$$\mathbf{z}^n = \operatorname*{argmin}_{\mathbf{z} \in S} \frac{1}{2\tilde{\alpha}} \|\mathbf{z} - (\mathbf{z}^n - \alpha D'(\mathbf{z}^n) - \alpha\beta R'(\mathbf{f}^n))\|_2^2 \quad (15)$$

in which $R'(\mathbf{f}^n) = \mathbf{f}^n - \mathbf{a}(\mathbf{f}^n)$, for some denoiser $\mathbf{a}$ (3). In the case where $D$ is assumed quadratic, (15) tells us the optimal soft target $\mathbf{z}^n$ is a combination of $(\mathbf{z}^n, \mathbf{f}^n, \mathbf{a}(\mathbf{f}^n))$ with weights $(1 - \alpha, \alpha(1 - \beta), \alpha\beta)$ and can be written as (8). The weights are non-negative for $0 \le \tilde{\beta} \le \lambda$, in which case $0 \le \alpha \le 1$ due to the upper bound on the step size $\tilde{\alpha}$ being $1/\lambda$. Any convex combination of $(\mathbf{z}^n, \mathbf{f}^n, \mathbf{a}(\mathbf{f}^n))$ belongs to $S$ and no projection is required. Due to this convenience, we use (8) by assuming the loss $D$ is locally quadratic even for a non-quadratic $D$.

As an example of a non-quadratic $D$, the Dice loss is used popularly in medical image reconstruction. We can write the gradient of the Dice loss (with respect to the first argument) as $D'(\mathbf{z}, \mathbf{f}) = \mathbf{C}^{-1}(\mathbf{D}\mathbf{z} - \mathbf{f})$, in which $\mathbf{D}$ is a diagonal matrix of class Dice scores, and $\mathbf{C}$, of average class pixel counts. More specifically, $\mathbf{D}$ and $\mathbf{C}$ are functions of $(\mathbf{z}, \mathbf{f})$ given by

$$d_{jj} = \frac{2\langle \mathbf{z}_j, \mathbf{f}_j \rangle}{\|\mathbf{z}_j\|_2^2 + \|\mathbf{f}_j\|_2^2}, \qquad c_{jj}^{-1} = \frac{2}{\|\mathbf{z}_j\|_2^2 + \|\mathbf{f}_j\|_2^2}, \quad (16)$$

respectively, where $\mathbf{u}_j$ denotes the $j$th probability map of $\mathbf{u}$.

In the case of the Dice loss $D$, one alternative to projected gradient descent (15) is exponential descent [80], where the quadratic term of (14) is replaced with the Kullback–Leibler

divergence $-\frac{1}{\alpha} D_{\mathrm{KL}}(\mathbf{z}, \mathbf{z}^n)$. This gives us the descent step

$$\mathbf{z}^n = \operatorname{softmax}(\log(\mathbf{z}^n) - \alpha D'(\mathbf{z}^n) - \alpha\beta R'(\mathbf{f}^n)), \quad (17)$$

guaranteeing $\mathbf{z}^n \in S$ by construction. With $N$ segmentation classes, convergence of this exponential descent step is better than the projected descent one by a factor of $O(\sqrt{N}/\sqrt{\ln N})$ suggesting that (17) be used over projected gradient descent especially for a large $N$. See [80] for further discussion.

## A.2    Deriving Temporal Ensembling

Temporal ensembling [23] can be interpreted as the update steps associated with optimization of the training objective

$$F(\Theta, \mathbf{z}^{\{n\}}) = G(\Theta) + \frac{1}{U} \sum_{n=1}^{U} (\lambda/2) \|\mathbf{z}^n - \mathbf{f}(\mathbf{u}^n | \Theta)\|_2^2, \quad (18)$$

that is, a degenerate case of (13) in which loss $D$ is quadratic and regularization term $\tilde{\beta} R = 0$. In this case, the minimizer of (14) is reducible to $\mathbf{z}^n = (1 - \alpha)\mathbf{z}^n + \alpha\mathbf{f}(\mathbf{u}^n | \Theta^n)$, which is the $\mathbf{z}$-update step of [23]. Since this $\mathbf{z}$-update is followed by a network weight update, too large a step size $\alpha$ can lead to sub-optimal training outcomes and even divergence. This is evidenced by the sub-optimal performance of the $\Pi$ model [58], which is a special case of temporal ensembling [23] that implicitly uses step size $\alpha = 1$. We stress that when $\mathbf{f}$ is non-negative and strictly convex in weights $\Theta$, the second term of objective $F$ has no regularizing effect and the optimal soft targets are given trivially by $\mathbf{z}^n = \mathbf{f}(\mathbf{u}^n | \Theta)$.

The Mean Teacher model of Tarvainen and Valpola [35] is a variant of temporal ensembling that generates soft targets using averaged weights rather than average reconstructions as in [23]. The training objective associated with this model can be expressed as

$$F(\Theta, \Omega) = G(\Theta) + \frac{1}{U} \sum_{n=1}^{U} (\lambda/2) \|\mathbf{f}(\mathbf{u}^n | \Omega) - \mathbf{f}(\mathbf{u}^n | \Theta)\|_2^2, \quad (19)$$

that is, we replace the $U$ many $\mathbf{z}^{\{n\}}$ optimization variables in objective (18) with parameters $\Omega$ of a "teacher" network. As before, one can alternately optimize over $\Theta$ and $\Omega$ with the other kept fixed. In particular, one can let $\mathbf{z}^n = \mathbf{f}(\mathbf{u}^n | \Omega)$ and use stochastic gradient descent similarly to (18) to update the parameters $\Theta$ of the "student" network.

Updating $\Omega^{n+1} = (1 - \alpha)\Omega^n + \alpha\Theta^n$ for teacher network parameters $\Omega$ can be seen as application of a quasi-Newton

TABLE B1
nn-UNet augmentation hyper-parameters for training method to the optimization of (19) over $\boldsymbol{\Omega}$. First, linearizing $\mathbf{f}(\mathbf{u}^n|\boldsymbol{\Omega})$ and $\mathbf{f}(\mathbf{u}^n|\boldsymbol{\Theta})$ in the respective parameters at $\boldsymbol{\Omega}^n$, we can write each loss term inside the summation of (19) as

$$D^n(\boldsymbol{\Omega}, \boldsymbol{\Theta}) \approx (\lambda/2)\|J\mathbf{f}(\mathbf{u}^n|\boldsymbol{\Omega}^n)(\boldsymbol{\Omega} - \boldsymbol{\Theta})\|_F^2, \quad (20)$$

neglecting all higher order terms, where $J\mathbf{f}(\mathbf{u}|\boldsymbol{\Omega})$ denotes the Jacobian (tensor) of $\mathbf{f}$ at $\boldsymbol{\Omega}$ for given $\mathbf{u}$. Taking a Newton step at the current iterate $\boldsymbol{\Omega}^n$ with step size $\alpha$ then corresponds to the update step for the teacher weights shown above.

| Parameter | Type | Range | Prob | Brain | Cortex |
|---|---|---|---|---|---|
| Random crop | shape | – | 0.00 | | |
| 2d/3d scaling | scale | $[0.70, 1.40]$ | 0.20 | ✓ | |
| Rand rotation | angle | $[-30°, 30°]$ | 0.20 | ✓ | |
| Rand flipping | – | – | 1.00 | ✓ | |
| Elastic, $16 \times 16$ | shift | $[-4.0, 4.0]$ | 1.00 | | ✓ |
| Gaussian noise | scale | $[0.00, 0.10]$ | 0.10 | ✓ | |
| Rand blurring | scale | $[0.50, 1.00]$ | 0.20 | | |
| xform low-res | scale | $[0.50, 1.00]$ | 0.25 | | |
| Scale intensity | scale | $[0.75, 1.25]$ | 0.15 | ✓ | |
| xform contrast | scale | $[0.75, 1.25]$ | 0.15 | ✓ | |
| Invert contrast | – | – | 0.10 | | |
| xform gamma | gamma | $[0.70, 1.50]$ | 0.30 | ✓ | |
| Conv (un)pool | – | – | – | | |
| Pooling layers | – | – | – | 4 | 4 |
| 2d/3d dropout | drop $p$ | – | – | 0.01 | 0.05 |

TABLE B2
Datasets used for the 3D brain reconstruction task.

| Split | ABD | ABD2 | ADH | GSP | MCIC | OAS | UKB | COB | PPMI | B39 | Total |
|-------|-----|------|-----|-----|------|-----|-----|-----|------|-----|-------|
| Train | – | – | – | – | – | – | – | – | – | 1 | 1 |
| Valid | 10 | 10 | 10 | 10 | 10 | 10 | 10 | 10 | 10 | 9 | 99 |
| Holdout | 55 | 55 | 55 | 55 | 55 | 55 | 55 | 50 | 55 | 10 | 500 |
| No label | 158 | 168 | 103 | 179 | 24 | 70 | 179 | 22 | 97 | 0 | 1000 |

TABLE B3
Dataset split used for the 2D cortical parcellation task.

| Dataset | Reconstructor Subjects | | | Denoiser Subjects | |
|---------|------|------|------|------|------|
| | Train | Val | Test | Train | Val |
| MindBoggle | 000–039 | 040–059 | 060–099 | 000–039 | 040–059 |
| HCP Images | 400–799 | 800–846 | 847–896 | – | – |
| HCP FS Labels | – | 800–846 | 847–896 | 000–399 | 800–846 |
| Total Subjects | 440 | 67 | 90 | 440 | 67 |

TABLE B4
Effect of training label set size on the denoiser accuracy.

| Training | 3D Brain Recon | | 2D Cortical Parcel | | Average | |
|----------|------|-------|------|-------|------|-------|
| | Dice | 95HD | Dice | 95HD | Dice | 95HD |
| 2 labels | 0.956 | 0.646 | 0.994 | 0.513 | 0.961 | 0.923 |
| 5 labels | 0.969 | 0.437 | 0.993 | 0.561 | 0.983 | 0.526 |
| 10 labels | 0.983 | 0.237 | 0.995 | 0.375 | 0.990 | 0.284 |
| 20 labels | 0.987 | 0.196 | 0.995 | 0.375 | 0.992 | 0.194 |
| 50 labels | **0.989** | **0.187** | **0.995** | **0.365** | **0.992** | **0.190** |
| Train data | 0.614 | 32.99 | 0.477 | 260.9 | 0.596 | 119.1 |

## Appendix B   Network Training Details

Here, we detail the training data and image augmentation strategies used for the reconstruction and dnoising tasks. In Table B1, we list the augmentation hyperparameters used in the anatomical brain segmentation and cortical parcellation tasks. Table B2 shows the breakdown of the 10 datasets used by the 3D brain reconstruction task for training, validation and testing. The label maps used to train the denoiser had no subject overlap with those used to train the reconsructor. In Table B3, we provide a breakdown of labeled (MindBoggle) and unlabeled (HCP) data for training the reconstructor and denoiser networks in the cortical parcellation task. Table B4 provides the accuracy of the denoiser trained using different numbers of training label maps.